# Heuristics for Routing and Spiral Run-time Task Mapping in NoC-based Heterogeneous MPSOCs

Mohammed kamel Benhaoua[1,2], Abbou El Hassen Benyamina[1] and Pierre Boulet[2]

[1] Department of Computer Science, Faculty of Sciences, University of Oran – Es Senia, Algeria
BP 1524, EL M'Naouer, Oran, Algeria

[2] University Lille 1, LIFL, CNRS, UMR 8022
F-59650 Villeneuve d'Ascq, France

**Abstract**
This paper describes a new Spiral Dynamic Task Mapping heuristic for mapping applications onto NoC-based Heterogeneous MPSoC. The heuristic proposed in this paper attempts to map the tasks of an applications that are most related to each other in spiral manner and to find the best possible path load that minimizes the communication overhead. In this context, we have realized a simulation environment for experimental evaluations to map applications with varying number of tasks onto an 8x8 NoC-based Heterogeneous MPSoCs platform, we demonstrate that the new mapping heuristics with the new modified dijkstra routing algorithm proposed are capable of reducing the total execution time and energy consumption of applications when compared to state-of the-art run-time mapping heuristics reported in the literature.

## 1. Introduction

Multiprocessor Systems-on-Chip (MPSoCs) is a solution that implements a multiple processing elements (PEs) in the same chip. Advancement in nanometer technology enables that a Future MPSoCs contained thousands of PEs in a single chip by 2015[3], [14], [8]. MPSoCs are being increasingly used in complex embedded applications. The Network-On-Chip (NoC) has been introduced as a power-efficient, scalable inter-communication, interconnection mechanism between PEs [14], [8]. Mapping is important phase in architectural exploration in NoC based MPSOC. The application and architectural platform are represented by processing model, application task graph and architectural graph respectively. Considering the moment when task mapping is executed, approaches can be either static or dynamic. Static mapping defines task placement at design time, having a global view of the MPSoC resources. As it is executed at design time, it may use complex algorithms to better explore the MPSoC resources, resulting in optimized solutions [24],[9],[16],[7],[12], [23],[19]. However, static mapping is not able to handle a dynamic workload, new tasks or applications loaded at run-time. To cope with this feature of actual MPSoCs, Dynamic (runtime) mapping techniques are required to map them onto the platform resources [4],[21],[13],[5],[6],[17],[15].

The main goal of this paper is to present a new Spiral Dynamic Task Mapping heuristic for run-time mapping applications. The presented heuristics are applied onto NoC-based Heterogeneous MPSoC platform. Two types of PEs are considered instruction set processors (ISPs) and Reconfigurable Areas (RA). Instructions set processors are used to execute software tasks and Reconfigurable Areas for hardware tasks. Heuristic also try to map the tasks of an application in a clustering region to reduce the communication overhead between the communicating tasks. The heuristic proposed in this paper attempts to map the tasks of an applications that are most related to each other in spiral manner and to find the best possible path load that minimizes the communication overhead with using a newly modified dijkstra routing algorithm proposed also in this paper. The new presented heuristic show significant performance improvements when compared to the latest run-time mapping heuristics reported in the literature. The performance metric includes execution time and energy consumption.

The rest of the paper is organized as follows. Section2 provides an overview of related work. Section 3 presents the MPSoC architecture. In Section 4, proposed mapping strategies along with the routing algorithm are presented. Experimental set-up and the results are presented in Section 5 with Section 6 concluding the paper and providing future directions.

## 2. Related Work

Mapping of tasks onto the MPSoC platform require finding the placement of tasks into the platform in view of some optimization criteria like reducing energy consumption, reducing total execution time and optimizing occupancy of channels. If the MPSoC platform is heterogeneous, then a task binding process is required before finding the placement for a task. The binding process involves defining a platform resource for each task type like instruction set processors for software tasks and FPGA tiles for hardware tasks. Task mapping is accomplished by static (design-time) or dynamic (run-time) mapping techniques [3].

2.1 Static mapping techniques

The most existing work's in litterature to solve the problem of mapping in the NOC platform are Static mapping. Static mapping defines task placement at design time, having a global view of the MPSoC resources and the tasks of applications (tasks graph). As it is executed at design time, it may use complex algorithms to better explore the MPSoC resources, resulting in optimized solutions. Heuristics like Genetic approach and exact methods like Tabu Search and stimulated annealing are presented in [24],[7], [12],[23],[19]. In [9], [16], energy-aware mapping algorithms are presented. These techniques find fixed placement of tasks at design-time with a well known computation and communication behavior. Related works classified for static mapping techniques are shown in Table 1. However, static mapping is not able to handle a dynamic workload, new tasks or applications loaded at runtime. To cope with this feature of actual MPSoCs, Dynamic (run-time) mapping techniques are required to map them onto the platform resources.

2.2 Dynamic mapping techniques

The challenge in the latest work's to solve the problem of mapping in the NoC-based heterogeneous MPSoCs are to present run-time mapping techniques for mapping application's tasks onto them . Wildermann et al. [22] evaluate the benefits of using a runtime mapping heuristic (communication and neighborhood cost functions), which allows decreasing the communication overhead.

Holzenspies et al. [20] investigate another run-time spatial mapping technique, considering streaming applications mapped onto heterogeneous MPSoCs, aiming on reducing the energy consumption imposed by such application behaviors. Schranzhofer et al. [2] suggest a dynamic strategy based on pre-computed template mappings (defined at design time), which are used to define newly arriving tasks to the PEs at run-time. Carvalho at al. [11] evaluate pros and cons of using dynamic mapping heuristics (e.g. path load and best neighbor), when compared to static ones (e.g. simulated annealing and Taboo search). Carvalho's approach was extended by Singh et al. [3], [1], employing a packing strategy, which minimizes the communication overhead in the same NoC-based MPSoC platform. Additionally, Singh's approach was improved to support multitask mapping onto the same PE. Different mapping heuristics were used to evaluate the performance. According to the Authors, the communication overhead of the whole system is reduced, decreasing the energy consumption. Faruque et al. [18] propose a decentralized agent-based mapping approach, targeting larger heterogeneous NoC-based MPSoCs ( 32x64 system is used as case study).

The most related works classified for Dynamic mapping techniques are shown in Table 2. Mapping heuristics Nearest Neighbor (NN) and Best Neighbor (BN) presented in Carvalho and Moraes [10] and two run-time mapping heuristics presented in Singh et al. [3] are taken for evaluation and performance comparison with our new proposed mapping heuristics.

Table 1: Related work classified for static mapping techniques

| REF | Tasks Mapping | Communications Mapping | Tasks Scheduling | Communications Scheduling | Voltage selection link | Voltahe Selection tasks | Objectives | Guarantee hard deadlines |
|---|---|---|---|---|---|---|---|---|
| Y.Zhang and al 2002 | List Scheduling | ----- | List Scheduling | ----- | ----- | ILP | Min Energy consumption | No |
| S.Kumar and al 2003 | Genetic Algorithm | ----- | ASAP/ALAP | ----- | ----- | ----- | Max Performance time | No |
| R.Marculescu and al 2004 | List Scheduling | Deterministic | List Scheduling | Deterministic | ----- | ----- | -Max performance time -Min Energy consumption | Yes |
| D. Shin and al 2004 | Genetic Algorithm | Genetic Algorithm | List Scheduling | ----- | List Scheduling | ----- | Min Energy consumption | Yes |
| M.Armin n and al 2007 | List Priority (TPL,PPL) | ----- | ----- | ----- | ----- | ----- | Min Energy consumption | No |
| C.Chou and al 2009 | Heuristic | Heuristic | ----- | Deterministic | ----- | ----- | Max Performance time | Yes |
| F.Vardi and al 2009 | List Priority (TPL1,TPL2 ,RTPL,PPL) | ----- | ----- | ----- | ----- | ----- | -Max performance time -Min Energy consumption | No |
| X.Wang and al 2009 | Heuristic | Deterministic | ----- | ----- | ----- | ----- | Min Energy consumption | Yes |

## 3. Heterogeneous MPSoC Architecture

MPSoC architecture used in this work contains a set of different processing elements which interact via a communication network [14]. Software tasks execute in instruction set rocessors (ISPs) and hardware tasks execute in reconfigurable logics (reconfigurable area-RA) or in dedicated IPs.

| Authors | Mono / Multi Tasks | Type of Architecture | Type of control | Optimizations Goals |
|---|---|---|---|---|
| Smit 2005 | Mono | heterogeneous | Centralised | Energie Consumption |
| Ngouanga 2006 | Mono | homogeneous | Centralised | Communication volume, Computational Load |
| Hölzenspies 2007/2008 | Mono | heterogeneous | Centralised | Energie Consumption |
| Chou 2007/2008 | Mono | homogeneous | Centralised | Energie Consumption |
| Al Faruque 2008 | Mono | heterogeneous | Distributed | Execution time, Mapping time |
| Mehran 2008 | Mono | homogeneous | Centralised | Mapping time, Energie Consumption |
| Wildermann 2009 | Mono | homogeneous | Centralised | communication, latency Energie Consumption |
| Schranzhofer 2010 | Mono | homogeneous | Centralised | Energie Consumption |
| Carvalho 2010 | Mono | heterogeneous | Centralised | Communication volume Network contention |
| Singh 2009, 2010 | Multi | heterogeneous | Centralised | Energie Consumption, Communication volume, Network contention |
| Marcelo 2011 | Multi | homogeneous | Centralised | Tasks inter-dependency evaluation, energy consumption |
| Proposed Work | Mono | heterogeneous | Centralised | Communication volume, Network contention, Execution time, energie consumption |

Table 2: Related work classified for Dynamic mapping techniques

One of the processing node is used as the Manager Processor (M) that is responsible for task scheduling, task binding, task placement (mapping), communication routing, resource control and reconfiguration control. The M knows only the initial tasks of the applications. The initial task of each application is started by the M and new communicating tasks are loaded into the MPSoC platform at run-time from the task memory when a communication to them is required and they are not already mapped.

## 4. Proposed Approach

This section describes our proposed approach. Firstly we describe our dynamic spiral task mapping. Secondly we describe the modified dijkstra routing algorithm. First we introduce some definitions for proper understanding of the proposed approach.

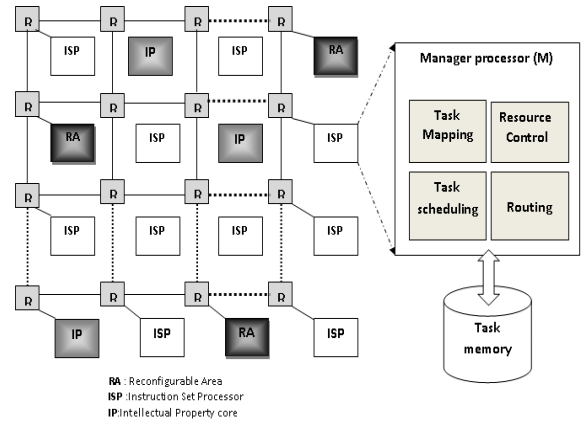

Fig. 1: Conceptual MPSOC architecture

### 4.1 Definitions:

Definition 1:

An application task graph is represented as an acyclic directed graph $TG = (T,E)$, where $T$ is set of all tasks of an application and $E$ is the set of all edges in the application. Figure 2 (a) describes an application having initial, software and hardware tasks along with the edges $(E)$ connecting these tasks and (b) shows the master-slave pair (communicating tasks). The starting task of an application is the initial task that has no master. $E$ contains all the pair of communicating tasks and is represented as $(mtid, stid, (V_{ms}, V_{sm}))$, where $mtid$ represents the master task identifier, $stid$ represents the slave task identifier; $V_{ms}$ is the data volumes from master to slave; $V_{sm}$ is the data volumes from slave to master.

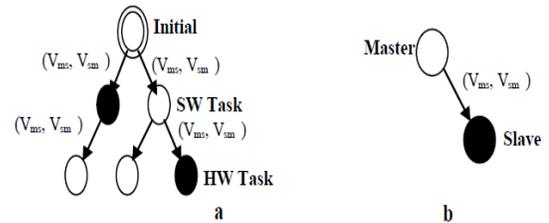

Fig. 2: Application modeling and Master-Slave

Definition 2:

A NoC-based heterogeneous MPSoC architecture is a directed graph $AG = (P, V)$, where $P$ is the set of tiles $p_i$ and $v_{i,j}$ presents the physical channel between two tiles pi and pj. A tile pi consists of a router, a network interface, a heterogeneous processing element, local memory and a cache.

Definition 3:

The application mapping is represented by mpng : ti→|pi to map the tasks of the application onto the NoC-based heterogeneous MPSoC.

4.2 Reference dynamic mappings heuristics

*1) The First Free (FF) heuristic:* Simply selects the next compatible processor to map a given task, thus walking sequentially through all processors before considering an processor again.

*2) Minimum Maximum Channel load (MMC) heuristic:* Considers all possible mappings for a given task and chooses the one that increases the least the peak load of a channel of the NoC.

*3) Minimum Average Channel load (MAC) heuristic:* Considers all possible mappings for a given task and chooses the one that increases the least the average load of the channels of the NoC

*4) The Nearest Neighbor (NN) heuristic:* Considers only the proximity of an available resource to execute a given task. NN starts searching for a free PE able to execute the target task near the source task. The search tests all n-hop neighbors, n varying between 1 and the NoC limits

*5) The Path Load (PL) heuristic :* Computes the load in each channel used in the communication path. PL computes the cost of the communication path between the source task and each one of the available resources. The selected mapping is the one with minimum cost.

*6) The Best Neighbor (BN) heuristic:* Combines NN search strategy with the PL computation approach. The search method of BN is similar to NN.

4.3 Proposed spiral heuristic based on our Modified Dijikstra routing Algorithm

*1) Spiral heuristic:* To Map the applications, firstly the initials tasks of applications are placed in distributive way the farest possible between them in a middle of the clusters, using a strategy of clusters like shown in Fig. 3. This permit the same tasks of application could be placed in a same region near between them, which reduces the communications costs. The frontiers of clusters are virtual and the common regions could be shared by the tasks of different applications.

After that the initials tasks of each application are placing communicative tasks ask to be placed. To place required task, the master processor (M) try to place it around the processor which has executed the appealed task going from a distance equal 1 (hop) until the limit of NoC. The ressource (the processor according to type of the task) is researched on spiral manner according to sequencement 1, 2, 3, 4, 5, 6, 7, 8 like shown in the figure Fig .3. Explores spiral neighbors and performs the mapping this prevents the calculation of all possible solutions mapping, as in the PL and calculation of the best neighbor as in the BN heuristic, reducing the execution time for mapping.

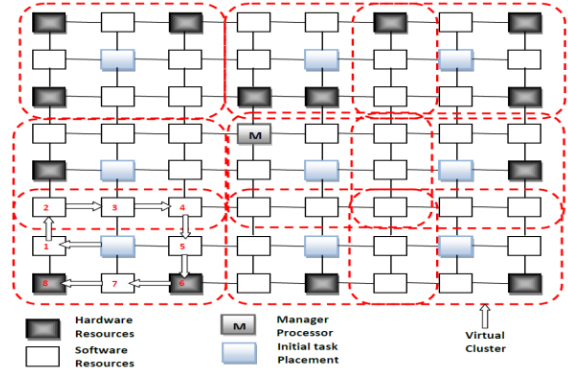

Fig. 3: Initial tasks placement for mapping applications with Spiral packing strategy.

*2) Modified dijkstra routing algorithm:* After mapping the communicating tasks, we need a communication mapping between them. Our new proposed communication mapping tries to search a best path with a high bandwidth. The proposed heuristic reduce the computational time and energy consumption.

```
Algorithm 1 SPIRAL HEURISTIC
Entrées:  NoFreeResources , CurrentProcessor, PE,
FreeResources,  HopDistance,  NocLimit  Sortie:
FreeElement
 1: HopDistance ← 0
 2: HopX ← 0
 3: HopY ← 0
 4: while (FreeElement = NULL) AND HopX != NocLimit
    AND HopY!= NocLimit do
 5:    if (CurrentProcessor.left.isFree()) then
 6:       FreeElement ← CurrentProcessor.left;
 7:    else
 8:       HopX- -;
 9:    end if
10:    if CurrentProcessor.top.isFree() then
11:       FreeElement ← CurrentProcessor.top;
12:    else
13:       HopY- -;
14:    end if
15:    if CurrentProcessor.right.isFree() then
16:       FreeElement ← CurrentProcessor.right;
17:    else
18:       HopX+ +;
19:    end if
20:    if CurrentProcessor.bottom.isFree() then
21:       FreeElement ← CurrentProcessor.bottom;
22:    else
23:       HopY+ +;
24:    end if
25: end while
```

```
Algorithm 2 MODIFIED DIJKSTRA ROUTING ALGORITHM
Entrées: IdSource , IdDestination Sortie: BestTraject
    /*While we did not reach the destination */
 2: while (stop=false) do
    /*Got back the neighbor who has the least used link*/
 4:    min ← minWeight();
    if min <> idDest then
 6:        /*Add the link in the list of the path*/
        Listpath.add(min)
 8:    else
        stop ← true
10:    end if
    end while
12: BestTraject ← Listpath
```

## 5. Experimental Set-Up and the Results

For the experimentation we have used a language of programming of high level which the JAVA language.

### 5.1 Experimental set-up

We have realize a heterogeneous platform simulated which comprises 64 processors of which 14 hardware, 49 software and a processor which is used like processor manager which is responsible of placement of the applications tasks, the configuration and update the platform, the communications routing. This platform uses a network on chip for the communication. We have used XML file for describing graphs of tasks used which are the same used in work of A.K.Singh, tasks (initial,software and hardware). The time processing of tasks depends on the specificity and the capacity of processor. we have fixed parameter software processors needs 40 cycles for an instruction, however hardware processors is fast and needs 20 cycles for one instruction. In a reverse of the consumption of energy or the processors hardware consume more than the processors software that we have fixed to 20 and 10 respectively. The shape of tasks is fixed to a number of instructions. The shape of data changed is 100 parquets. The used scenario is a number of one, three, seven and ten (10) applications witch proceeds between 7 to 9 tasks.

The platform is divided to nine (9) clusters which permits to launch nine (9) applications en parallel beyond this number the others applications which requires to be placed, have to waited in queue. For the placement of the tasks of applications we have implemented our proposed Dynamic task mapping based on Spiral packing strategy and Modified Dijkstra routing method. The spiral method tries to map the tasks in close manner with minimum exploration of the NoC space. The implementation of our method Modified Dijkstra routing which minimize the time processing and energy consumption of the system. For a comparison study we have implemented the NN and BN dynamic mapping heuristics.

### 5.2 Experimental results

We have executed the implemented dynamic heuristics the Nearest Neighbor (NN) and the Best Neighbor (BN) for the placement of one, three, seven and ten applications in parallel on the platform simulated of 64 processors whose 14 hardware, 49 software and one for the processor manager (M). For the same we have executed our proposed Heuristics for Routing and Spiral Run-time task Mapping. For the measurements of performances we have calculated the execution time and the energy consumption. The Fig.4 show the optimization brought by our approach in terms of execution time awards the use of the proposed approach. The Fig.5 show the optimization brought by our approach in terms of energy consumption towards the use of the proposed approach.

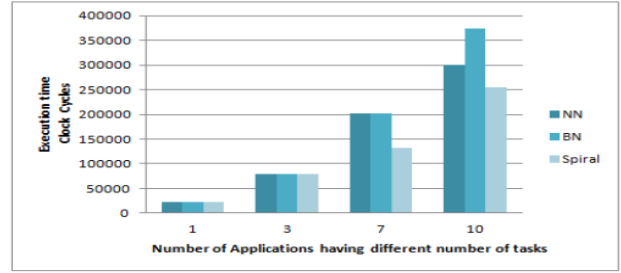

Fig. 4: Execution Time comparison of proposed approach with NN and BN respectively

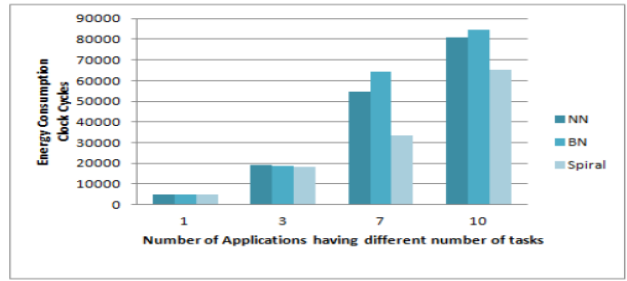

Fig. 5: Energy Consumption comparison of proposed approach with NN and BN respectively

## 6. Conclusion and Future Directions

A new dynamic task mapping heuristic that try to map the application tasks in close manner to reduce the cost of communications is proposed. To reduce more optimal the cost of the communications a Modified Dijkstra routing is proposed also in this paper. Through the environment of simulation that we have realized the comparative study between the nearest neighbor (NN), the best neighbor (BN) and the proposed dynamic tasks mapping heuristic based on the spiral packing strategy and Modified Dijkstra routing Algorithm. We have showed the offered optimization by our approach. In our future research works the right challenge is trying to propose others strategies of research other than the one used in the latest works and Proposition of other heuristics of dynamic mapping.

**Benhaoua Mohammed Kamel :** Received his Magister degree in computer science, Oran University, Algeria, in 2009. Currently, he is student towards the completion of his PhD. His research interests include NoC-based MPSoC design, run-time mapping algorithms.

**Benyamina Abbou el hassen :** received his Ph.D. degree in Computer Science in 2008 from University of Oran (Algeria), He is assistant professor at the university (Algeria). His research works include parallel processing, optimization, design space exploration and Model Driven Engineering with the special focus on real-time and embedded systems.

**Pierre Boulet :** received his Ph.D. degree in Computer Science in 1996 from University of Lyon, He was assistant professor at the university Lille 1 from September 1998 to August 2002, and then became a researcher at INRIA Futurs from September 2002 to August 2003. Since September 2003, he is a full professor at the university Lille 1. His research works include parallel processing, optimization, design space exploration and Model Driven Engineering with the special focus on real-time and embedded systems.